\documentclass[twocolumn,nofootinbib,10pt,prb,floatfix]{revtex4-1}
\usepackage[T1]{fontenc}
\usepackage[utf8]{inputenc}
\usepackage[english]{babel}
\usepackage{amsmath}  
\usepackage{amsfonts} 
\usepackage{graphicx} 
\usepackage[pdftex,bookmarks=true,bookmarksopen,bookmarksnumbered,colorlinks,linkcolor=blue,citecolor=blue]{hyperref}
\graphicspath{ {./} }
\usepackage{natbib}
\usepackage[table]{xcolor}
\usepackage{ulem}
\usepackage{xcolor}
\usepackage{braket}
\usepackage{booktabs}
\usepackage{multirow}
\usepackage{caption}
\usepackage{float} 
\usepackage{upgreek}
\usepackage[modulo,pagewise]{lineno}
\newcommand{\aref}[1]{\hyperref[#1]{Appendix~\ref*{#1}}}
\makeatletter
\def\blfootnote{\gdef\@thefnmark{}\@footnotetext}
\makeatother

\begin{document}
\captionsetup[table]{name={Table},labelsep=period,justification=raggedright,font=small}
\captionsetup[figure]{name={Fig.},labelsep=period,justification=raggedright,font=small}
\renewcommand{\equationautorefname}{Eq.}
\renewcommand{\figureautorefname}{Fig.}
\renewcommand*{\sectionautorefname}{Sec.}

\title{Entangling gates on degenerate spin qubits dressed by a global field}
\author{Ingvild Hansen$^{1*}$}
\author{Amanda E. Seedhouse$^{1,2}$}
\author{Santiago Serrano$^{1}$}
\author{Andreas Nickl$^1$}
\author{MengKe Feng$^{1,2}$}
\author{Jonathan Y. Huang$^{1}$}
\author{Tuomo Tanttu$^{1,2}$}
\author{Nard Dumoulin Stuyck$^{1,2}$}
\author{Wee Han Lim$^{1,2}$}
\author{Fay E. Hudson$^{1,2}$}
\author{Kohei M. Itoh$^{3}$}
\author{Andre Saraiva$^{1,2}$}
\author{Arne Laucht$^{1,2}$}
\author{Andrew S. Dzurak$^{1,2\dagger}$}
\author{Chih Hwan Yang$^{1,2\ddagger}$}
\affiliation{$^1$School of Electrical Engineering and Telecommunications, The University of New South Wales, Sydney, NSW 2052, Australia}
\affiliation{$^2$Diraq, Sydney, New South Wales, Australia}
\affiliation{$^3$School of Fundamental Science and Technology, Keio University, Yokohama, Japan}

\date{\today}
\begin{abstract}
Semiconductor spin qubits represent a promising platform for future large-scale quantum computers owing to their excellent qubit performance \cite{Xue2022,Noiri2022,madzik2022}, as well as the ability to leverage the mature semiconductor manufacturing industry for scaling up \cite{zwanenburg2013,zwerver2022,Gonzalez2021}. Individual qubit control, however, commonly relies on spectral selectivity \cite{Veldhorst2014,fogarty2018integrated,takeda2016fault}, where individual microwave signals of distinct frequencies are used to address each qubit. As quantum processors scale up, this approach will suffer from frequency crowding, control signal interference and unfeasible bandwidth requirements. Here, we propose a strategy based on arrays of degenerate spins coherently dressed \cite{mollow1969,xu2007,Baur2009,Londonn2013} by a global control field \cite{hansen2022,Seedhouse2021} and individually addressed by local electrodes. We demonstrate simultaneous on-resonance driving of two degenerate qubits using a global field while retaining addressability for qubits with equal Larmor frequencies. Furthermore, we implement SWAP oscillations during on-resonance driving, constituting the demonstration of driven two-qubit gates. {Significantly, our findings highlight how dressing can overcome the fragility of entangling gates between superposition states \cite{Petit2022,Petta2005,Simmons2019,Guo2023} and increase their noise robustness.} These results constitute a paradigm shift in qubit control in order to overcome frequency crowding in large-scale quantum computing. 
\end{abstract}

\maketitle
In the race towards building a large-scale universal quantum computer, we are faced with several potential future bottlenecks. The fragility of qubits is a commonly discussed aspect as current quantum error correction codes, essential for fault tolerant quantum computing, only allow for very small error rates \cite{fowler2012}. The scaling prospect is also of utmost importance, since the number of required physical qubits is expected to exceed millions. Challenges include routing of necessary control signals onto the quantum processor chip \cite{franke2019rent}, control signal interference, variability, \cite{Cifuentes2023} etc. Scaling up current architectures by brute force will present many challenges and is not necessarily the best course of action.\blfootnote{$^*$ \textcolor{blue}{i.hansen@unsw.edu.au}}\blfootnote{$^{\dagger}$ \textcolor{blue}{a.dzurak@unsw.edu.au}}\blfootnote{$^{\ddagger}$ \textcolor{blue}{henry.yang@unsw.edu.au}}

Global control was suggested by Kane in 1998 \cite{kane1998} and involves globally applying a single microwave field to an array of qubits. Recent advancements have demonstrated a version where the qubits are by-default on resonance \cite{Seedhouse2021,hansen2022,Hansen2021}. Dressed qubits are continuously decoupled from low-frequency noise and individually addressed by local electrodes \cite{laucht2017dressed,hansen2022}. Furthermore, the global field can be generated off-chip \cite{vahapoglu,vahapoglu2022}, which frees up space on the chip and simplifies control signal routing. Therefore, a global control scheme based on dressed degenerate qubits tackles the fragility of qubits while offering a prospect for scalability. Additional advantages include reduced control bandwidths and control signal interference.

In this work, we demonstrate two crucial components of the vision to operate dressed degenerate qubits in a global field, namely single-qubit addressability and two-qubit operation. We tune two silicon metal-oxide-semiconductor (SiMOS) quantum dot spin qubits (see Extended data \autoref{fig:ext0}) such that their Larmor and Rabi frequencies are matched, and then perform single- and two-qubit universal dressed control in a global field. 

\begin{figure*}[hbt!]
    \centering
    \includegraphics[width=1\textwidth]{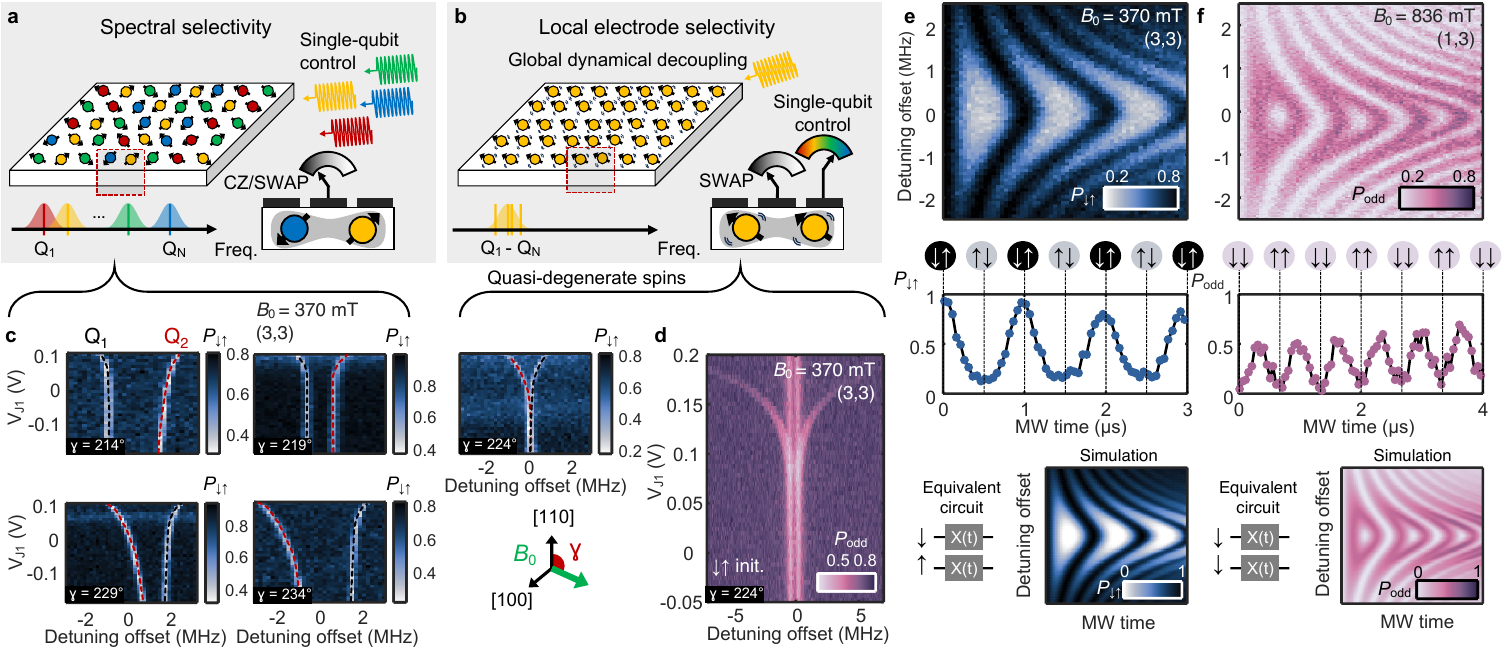}
    \caption{{\bfseries{Degenerate spin qubits in a global field. }}{Operation using (a) spectral selectivity of non-degenerate spins and (b) local electrode selectivity of degenerate spins for individual qubit control.} (c) Adiabatic inversions of two spins with varying static magnetic field angle $\gamma$ using $\ket{\downarrow\uparrow}$ initialisation and adiabatic singlet-triplet (ST) readout. The frequency detuning is offset for each sub-panel. (d) High-resolution adiabatic inversion with parity readout at $\gamma=224^{\circ}$ where the spin qubits are quasi-degenerate. (e,f) Rabi chevrons of quasi-degenerate spins simultaneously driven by a global field with adiabatic ST and parity readout, respectively. The initial states are $\ket{\downarrow\uparrow}$ and $\ket{\downarrow\downarrow}$, respectively, with Larmor frequencies $10.362$\,GHz and $23.221$\,GHz. Traces at the centre frequencies are plotted below. Circuit models and simulations are shown in the bottom of panels (e-f). 
}
    \label{fig:fig1}
\end{figure*}

\subsection*{Degenerate spins in a global dressing field}
\label{sec:req}
Traditionally, qubits are addressed by their individual Larmor frequencies using spectral selectivity \cite{Veldhorst2014,fogarty2018integrated,takeda2016fault}, {as illustrated in \autoref{fig:fig1}(a).} To avoid interference between different control signals, this strategy requires a spread of Larmor frequencies in excess of the Rabi frequency. The Larmor frequency of a qubit is set by the Bohr magneton, the static magnetic field strength $B_0$ and the $g$-factor. The latter has a natural variability in Si/SiO$_2$ devices ($\sim$ MHz$/$T) due to spin-orbit coupling \cite{Veldhorst2014,tanttu2019,harvey2019spin,Cifuentes2023,Martinez2022}. Nanomagnets resulting in a slanted field can also be used such that qubits in different locations see different magnetic field strengths  \cite{Tokura2006,philips2022,Takeda2021,pioro2008electrically}. With these aforementioned strategies one can control a handful of qubits \cite{philips2022,Takeda2021,Lawrie2023,Hendrickx2021,Lawrie2020}. However, when scaling up to millions of qubits, frequency crowding will become a problem \cite{Jones2018,Seedhouse2021}. {Moreover, due to the distribution of Larmor frequencies, the native two-qubit gate between neighbouring spins, which is a function of the Larmor frequency difference and the exchange magnitude, varies between controlled-Z (CZ) and SWAP \cite{Meunier2011}.}

The aim of a global control strategy using dressed qubits is to have arrays of {quasi-degenerate} spins that can be driven on-resonance by a global microwave field and addressed individually by local electrodes, {as illustrated in \autoref{fig:fig1}(b)}. This approach circumvents the problem of frequency crowding and control signal interference when scaling up. {Conveniently, the SWAP gate is the native two-qubit gate for quasi-degenerate spins due to the small Larmor frequency differences \cite{Meunier2011}.}

To drive multiple qubits with a global field, it is favourable for the qubits to be uniform, and it has been shown recently that the variability of SiMOS spin qubits caused by the roughness of the Si/SiO$_2$ interface is bounded \cite{Cifuentes2023}. The $g$-factor variability is expected to be $<0.1$\,\% for a DC magnetic field applied along the [100] direction and the robust control protocol introduced in Refs.~\onlinecite{Hansen2021,hansen2022} and employed in Ref.~\onlinecite{vallabhapurapu2023high} is designed to handle this bounded variability. 

Here, we mimic the behaviour of a global control field with an electron-spin-resonance (ESR) antenna and therefore have to carefully match the Larmor and Rabi frequencies. The qubit Larmor frequencies must all be within the linewidth of the global microwave field in order to achieve on-resonance driving and the Rabi frequencies must be similar, so that the qubits follow the same clock cycle. A sizeable Stark shift from the top gates on the Larmor frequencies is also important for addressability.  

\subsection*{Larmor frequency matching}
\label{sec:Larmor}
 The $g$-factor variability from spin-orbit interactions can be minimised by pointing the static magnetic field along $[100]$ instead of the default direction $[110]$ \cite{tanttu2019,Cifuentes2023}. The $[100]$ magnetic field orientation can also reduce the qubit susceptibility to charge noise and potentially increase $T_2^*$ \cite{tanttu2019,Ferdous2018}. Although this strategy is useful for matching the Larmor frequencies of two qubits, one can not rely on this alone when scaling up to larger numbers of qubits. Using low static magnetic field strength is also beneficial for the qubit uniformity \cite{fogarty2018integrated}, as the Larmor frequency variability is directly proportional to the $g$-factor times the static magnetic field ($\nu\propto g{B_0}$). Applying large microwave powers increases the Rabi frequency, and therefore also increases the tolerable spread of qubit Larmor frequencies. 

By ramping the voltage detuning between quantum dots at a determined rate, we initialise the system in the $\ket{\downarrow\uparrow}$ spin state, facilitated by the fact that near the charge transition the spins are not yet entirely degenerate. We then study the voltage dependence of the qubit frequencies by performing slow chirped microwave pulses, which adiabatically invert the spin state if its resonance frequency is contained within the range of the chirp~\cite{lauchtadb2014}. This is verified by measuring the probability that the final spin state is blockaded, as plotted in \autoref{fig:fig1}(c). 

Adiabatic spin inversions as a function of V$_{\rm{J1}}$ with different static magnetic field angles $\gamma$ are shown. At $\gamma=224^{\circ}$, near the $[100]$ orientation, the two qubits are quasi-degenerate. Note that after this angle, the order of the qubit frequencies inverts and therefore the initialisation also inverts, resulting in the reflection of the branching of the frequencies as a function of V$_{\rm{J1}}$. In \autoref{fig:fig1}(d) a high-resolution version of the quasi-degenerate case is shown with Larmor frequencies around 10.362\,GHz, separated by $<200$ kHz. This small difference in Larmor frequency is achieved by pointing the static magnetic field along [100], which minimises the Dresselhaus spin-orbit coupling, and using a relatively low magnetic field strength of 370\,mT.

The difference between spin Zeeman energies, combined with other parameters, is known to shift the outcome of the spin blockade from a singlet-triplet (ST) readout to a parity readout~\cite{Seedhouse2021_0}. In the present case, the $\ket{\downarrow\uparrow}$ state is mapped into the unblockaded singlet through an adiabatic ramp in some configurations (which we refer to as adiabatic ST), while in others any odd-parity state ends up equally unblockaded (see Supplementary  material). Once the operational parameters are settled, a characterisation is required to distinguish between the two scenarios. 
\begin{figure*}[hbt!]
    \centering
    \includegraphics[width=0.8\textwidth]{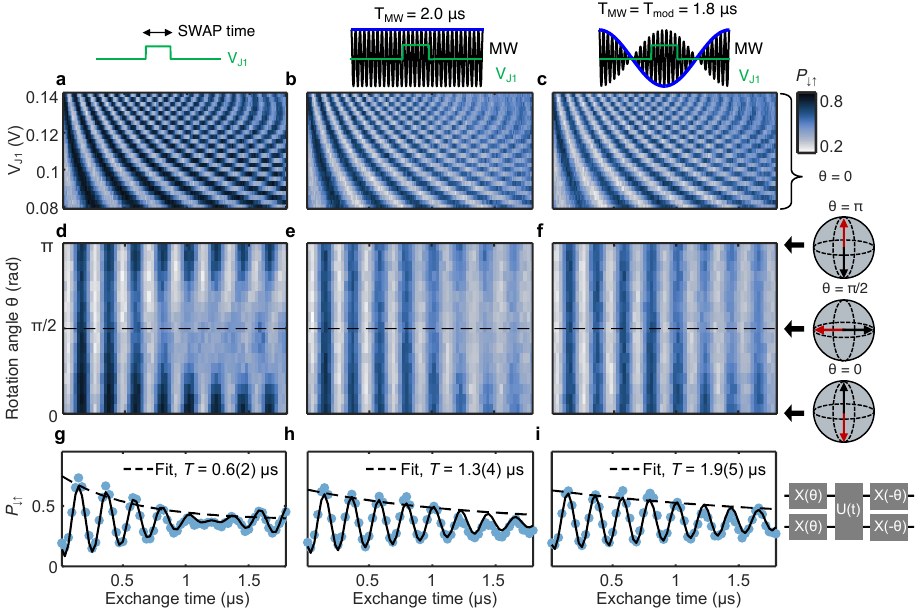}
        \caption{{\bfseries{Entanglement between on-resonance driven arbitrary spin states.}} SWAP oscillations during (a) no microwave drive, (b) continuous wave microwave of duration $T_{\rm{MW}}=2.0\,$\textmu{s} with $\Omega_{\rm{R}}=1\,$MHz, and (c) sinusoidal modulated microwave of duration $T_{\rm{MW}}=T_{\rm{mod}}=1.8\,$\textmu{s}. The microwave is applied along the $x$-axis and the modulation is according to $\sqrt{2}\Omega_{\rm{R}}\cos\big(2\pi{T_{\rm{mod}}}^{-1}t\big)$. (d-f) Initialisation of arbitrary states between $\ket{\downarrow\uparrow}$ and $\ket{\uparrow\downarrow}$ where $\theta$ is the rotation angle of the projection pulses and V$_{\rm{J1}}=0.1$\,V. In (g-i) the traces at $\theta=\pi/2$ ($\ket{y}\otimes\ket{\bar{y}}$ initialisation) are plotted and fitted, the solid line fits the theorised spin dynamics (see Supplementary material) and dashed line fits an exponential decay of the in-phase SWAP oscillations. All data is acquired in an interleaved manner. At zero time $P_{\downarrow\uparrow}\neq1$ for all V$_{\rm{J1}}$, despite $\ket{\downarrow\uparrow}$ initialisation, due to bandwidth limitations of the cables causing exchange interaction during the ramp.
    }
    \label{fig:fig2}
\end{figure*}
\subsection*{Rabi frequency matching}
\label{sec:Rabi}
When the Larmor frequencies of two qubits match, as demonstrated at low V$_{\rm{J1}}$ in \autoref{fig:fig1}(d), they can be driven simultaneously with a global field. The resulting Rabi frequency is set by the power delivered to the qubits from the microwave source, through the on-chip ESR antenna. Due to the transmission characteristics of the ESR antenna, the power delivered into the device varies as a function of microwave frequency (and therefore the Rabi frequency depends on the choice of $B_0$). Moreover, the Rabi frequencies of the two qubits are not necessarily equal. Besides the inhomogeneity of the oscillatory magnetic field generated by the antenna geometry, both qubits can be affected by the spurious electric component of the microwave field \cite{Gilbert2023}. This electric drive is usually minimised through the choices of double-dot electrostatic potential confinement, static magnetic field strength and orientation.

The only scalable way to address this variability in Rabi frequency is to engineer the microwave field to have minimal electric component. In the particular case of two qubits, however, we can tolerate the stray electric fields by fine-tuning the magnetic field strength until the Rabi frequencies of both qubits match. We find both Larmor frequency matching and Rabi frequency ($\Omega_{\rm{R}}$) matching with values 10.362\,GHz and 1\,MHz, respectively, at 370\,mT. The Larmor frequencies are different by $<20\%$ of $\Omega_{\rm{R}}$ and the Rabi frequencies by $<5\%$ of $\Omega_{\rm{R}}$. These conditions lead to qubit rotations that are sufficiently synchronous to consider the two qubits simultaneously driven with a shared clock (and any deviations are considered coherent errors in the qubit operations).

In \autoref{fig:fig1}(e) a Rabi chevron is shown for two qubits with matched Larmor and Rabi frequencies in the adiabatic ST readout regime with $\ket{\downarrow\uparrow}$ initialisation. The same is shown in \autoref{fig:fig1}(f) for the parity readout regime with $\ket{\downarrow\downarrow}$ initialisation. The adiabatic ST readout yields a periodic oscillation at the Rabi frequency but it is not a simple sinusoidal because only the $\ket{\downarrow\uparrow}$ is unblockaded, while the driving populates all states. The parity readout regime gives oscillations at twice the frequency and half the amplitude when initialised in $\ket{\downarrow\downarrow}$, due to $\ket{\downarrow\uparrow}$ and $\ket{\uparrow\downarrow}$ both being unblockaded (see Supplementary material). All the following measurements are done using $\ket{\downarrow\uparrow}$ initialisation and the adiabatic ST readout regime.

{By leveraging advanced microwave engineering and a dielectric resonator or a cavity as the microwave source in the future, increasing the magnitude and uniformity of the magnetic field, the matching conditions for Larmor and Rabi are not expected to present a significant challenge.}
\begin{figure*}[hbt!]
    \centering
    \includegraphics[width=\textwidth]{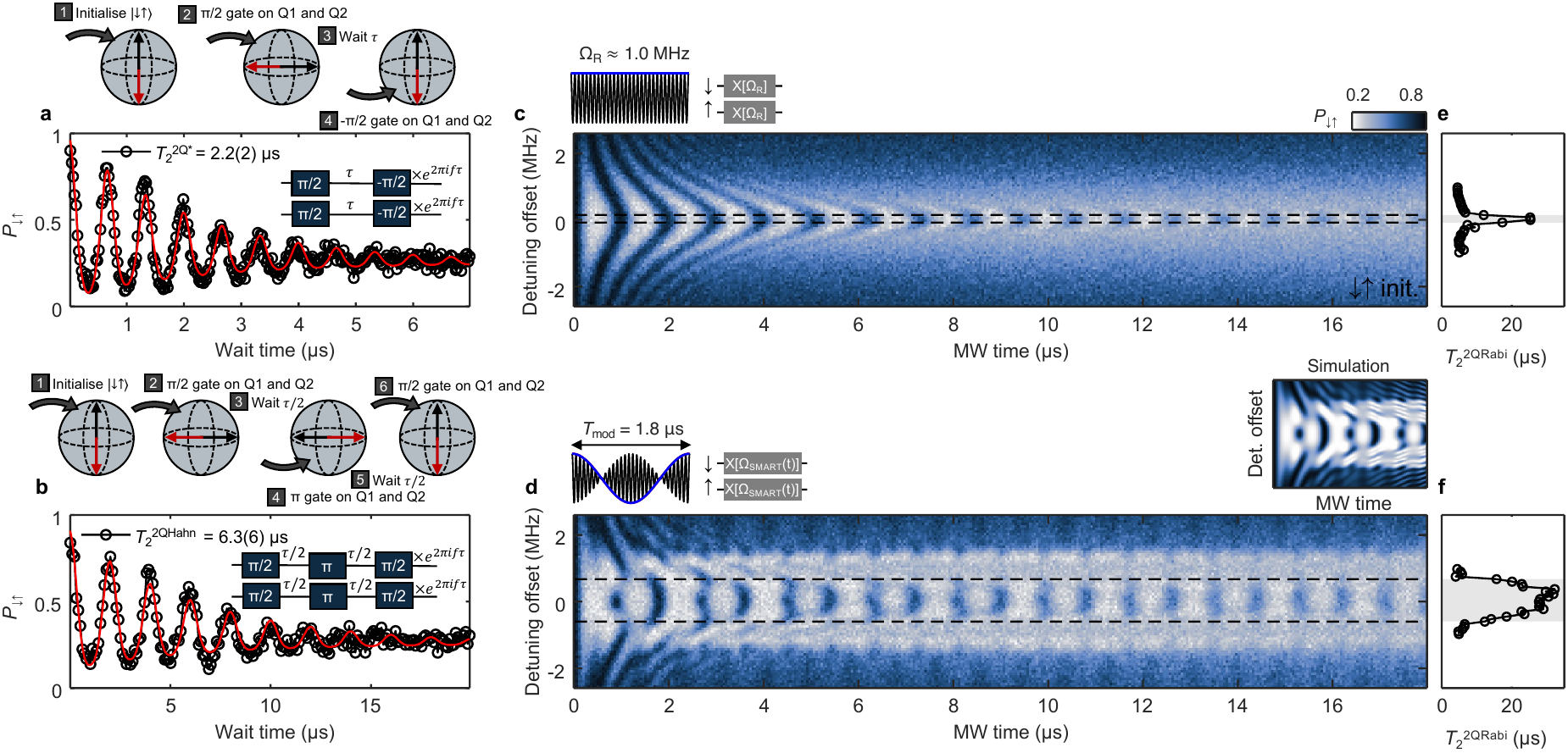}
    \caption{{\bfseries{Coherence metrics of degenerate qubits.}} (a) Free-induction decay and (b) Hahn echo measurement applied to degenerate qubits, where oscillations of frequency $f$ are introduced for fitting purposes. Chevron with (c) square modulation and (d) sinusoidal modulation acquired in an interleaved manner. Here, $\Omega_{\rm{SMART}}(t)=\sqrt{2}\Omega_{\rm{R}}\cos\big({2\pi{T_{\rm{mod}}^{-1}}}t\big)$. The extracted $T_2^{\rm{2QRabi}}$ times are plotted in (e-f) and the grey shaded area represents $T_2^{\rm{2QRabi}}>10\,$\textmu{s}.}
    \label{fig:fig3}
\end{figure*}
\subsection*{Entanglement during on-resonance driving}
\label{sec:2q}
The fundamental interaction between spins is the spherically symmetric Heisenberg exchange coupling $H_X= J \mathbf{\sigma_1} \cdot \mathbf{\sigma_2}$. In the particular case of non-degenerate spin qubits, the Larmor frequency difference creates an effective Ising coupling $H_I= J \sigma_{z,1} \sigma_{z,2}$, which favours the implementation of CZ gates. In the case of degenerate qubits the full Heisenberg interaction naturally leads to SWAP gates instead, as originally proposed by Loss and DiVincenzo \cite{Meunier2011,Seedhouse2021,Loss1998,Petta2005,Simmons2019,Petit2022}. 

A voltage pulse applied to gate J1 (see Extended data \autoref{fig:ext0}) turns the exchange interaction on and off, showing SWAP oscillations over time in \autoref{fig:fig2}(a) as a function of V$_{\rm{J1}}$. The fastest SWAP oscillation recorded here is $\sim100$\,ns, limited by the voltage range of the equipment (see Methods).  Differently from the case of CZ, where a strong exchange coupling leads to deviations from the idealised Ising model, here the stronger the exchange further accentuates the Heisenberg interaction. 

We repeat the same entangling pulse measurement in the case when a microwave on resonance with both qubits is applied with either a constant power or with a sinusoidal modulation, shown in \autoref{fig:fig2}(b-c) respectively. Note that the V$_{\rm{J1}}$ pulse is centred within the microwave pulse. The microwave pulse duration is fixed to ensure that it effectively performs an identity gate. We find that the SWAP oscillations remain unchanged during microwave driving but lose some visibility, which we attribute to degradation of the SET charge readout caused by heating. The sinusoidal modulated driving is found to be more robust against Larmor frequency variability (see Extended data \autoref{fig:ext2}). 

In \autoref{fig:fig2}(d-f) we showcase the most important result in this work, that the SWAP oscillations between superposition states \cite{Sigillito2019} are significantly more preserved for driven qubit implementations than for bare qubits. Looking at the experiment in \autoref{fig:fig2}(a-c) this point is missed because the SWAP oscillations between the states $\ket{\uparrow\downarrow}$ and $\ket{\downarrow \uparrow}$ are not exposed to $T_2$ decoherence times the whole time. However, SWAP oscillations between the more relevant $\left(\ket{\uparrow\downarrow} \pm \ket{\downarrow \uparrow}\right)/\sqrt{2}$ states reveal the true impact of decoherence on these states and the benefits of dynamical decoupling.

We repeat the measurement with initial states varying continuously from $\ket{\downarrow\uparrow}$ ($\theta=0$) to $\ket{\uparrow\downarrow}$ ($\theta=\pi$) with a microwave pulse to rotate the qubits before a fixed V$_{\rm{J1}}$ pulse and undo the rotation at the end. For initialisations around $\theta=\pi/2$, the driven SWAP oscillations remain coherent for longer than in the undriven case. This can be explained by noise decoupling \cite{Watson2018, Xue2022,Huang2023} resulting from the driving field \cite{hansen2022}. Crucially, our results draw attention to the fragility of entangling gates between arbitrary superposition states (as opposed to eigenstates, see Extended data \autoref{fig:ext}), which are typically not studied in literature \cite{Guo2023,Petit2022,Petta2005,Simmons2019}. In \autoref{fig:fig2}(g-i), the oscillations for initialisation in $\ket{y}\otimes\ket{\bar{y}}$ [dotted line at $\theta=\pi/2$ in \autoref{fig:fig2}(d-f)] are compared. Being able to perform SWAP/$\sqrt{\rm{SWAP}}$ on arbitrary states with high fidelity is not only important in universal quantum computing for implementing two-qubit computational gates, but also in the context of coherent quantum state transfer \cite{Jones2018}.

\subsection*{Joint coherence metrics}
\label{sec:ctlfr}
In \autoref{fig:fig3}(a-b) free-induction decay and Hahn echo measurements are performed. In the context of simultaneously driven qubits, however, these experiments must be reinterpreted. The joint probabilities $P_{\downarrow\uparrow}$ oscillate [similar to \autoref{fig:fig1}(e)] because the recovery $\pi/2$ gate is applied at an angle increasing with wait time $\tau$ in regard to the preparation gate (these oscillations are introduced to improve the fitting accuracy, see Supplementary material). The decay of these oscillations is the joint decay rate of the two spins, which are simultaneously prepared in superposition states $\ket{+y}=1/\sqrt{2}(\ket{\uparrow}+i\ket{\downarrow})$ and  $\ket{-y}=1/\sqrt{2}(\ket{\uparrow}-i\ket{\downarrow})$. 
If the decoherence processes are completely independent, both the free-induction and Hahn echo decays are expected to be twice as fast as the conventional single-qubit measurements. We extract $T_2^{\rm{2Q}*}=2.2(2)\,$\textmu{s} and $T_2^{\rm{2QHahn}}=6.3(6)\,$\textmu{s}, where the superscript 2Q refers to the fact that both qubits are involved in the dephasing. 

In \autoref{fig:fig3}(c-d) we show the driven two-qubit oscillations over time as a function of the microwave frequency in the case of optimally matched Larmor and Rabi frequencies. The constant power case [\autoref{fig:fig3}(c)] is the same experiment as shown before in  \autoref{fig:fig1}(e), but now with sufficiently long times to observe the decay of the driven qubits. The horizontal dashed lines indicate the range of detunings that yield $T_2^{2Q\rm{Rabi}}>10$\,\textmu{s}. This arbitrary reference line gives an indication of the tolerance of this type of driven qubit to uncertainties in the qubit frequency, which in this case is approximately 230\,kHz.

In \autoref{fig:fig3}(d) the microwave is amplitude-modulated  with a sinusoid according to Ref.~\onlinecite{Hansen2021}, using the same average microwave power as in \autoref{fig:fig3}(c). The period of the sinusoidal modulation is set to the theoretical optimal $T_{\rm{mod}}=1.8\,$\textmu{s}, which is determined by the Rabi frequency obtained at this set power \cite{hansen2022} (see Supplementary materials). Comparing the range between the dashed lines in this modulated case (approximately 1.27\,MHz) we can see a five-fold improvement in tolerance for the qubit detuning. This is further emphasized in the fitted Rabi coherence times plotted in \autoref{fig:fig3}(e-f). 

The sinusoidal modulation improves the tolerance of the architecture to the variability in Larmor frequencies for large arrays of spin qubits and also offers protection against microwave amplitude errors. This was predicted from theory in Ref.~\onlinecite{Hansen2021}.
\begin{figure*}[hbt!]
    \centering
    \includegraphics[width=1\textwidth]{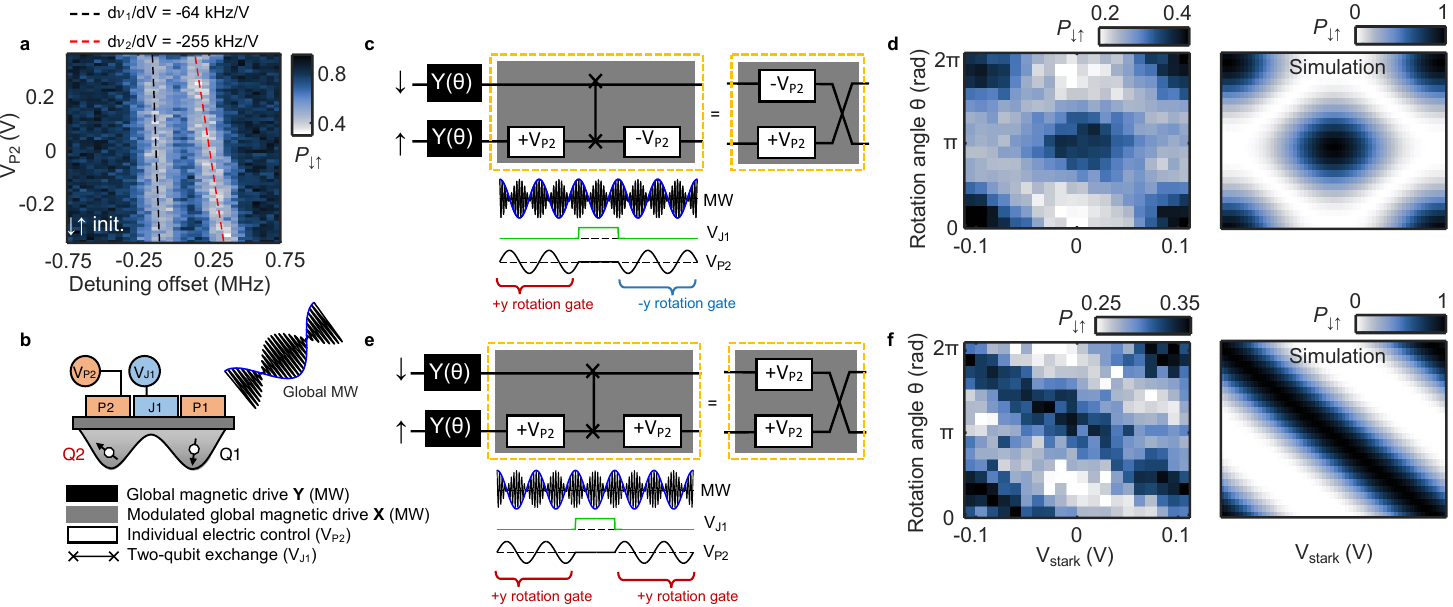}
    \caption{{\bfseries{Addressability of quasi-degenerate spins.}} (a) Stark shifts of the Larmor frequencies $\nu_1$ and $\nu_2$ with the voltage on gate P2. (b) Schematic of double dot and the relevant control knobs P2, J1 and global microwave. (c) Gate sequence implementing single-qubit $y$-rotations of both qubits under gate P2 by using a SWAP gate. The gate on Q1 and Q2 have opposite signs. A simplified gate sequence is also shown. (d) Experimental results and simulations for the sequence in (c). (e-f) Identical sequence as in (c-d) but with equal sign on the single-qubit gates. The single-qubit gates last for six periods of the global field [two periods shown in schematic (c,e)] and the SWAP gate for one period.
    } 
    \label{fig:fig4}
\end{figure*}
\subsection*{Addressability of quasi-degenerate spins}
\label{sec:Addressability}
Individual control of driven qubits is achieved by dynamically  shifting the qubit frequency leveraging the Stark effect created by the voltage bias at the top gate that forms the dot~\cite{laucht2017dressed,Hansen2021,hansen2022}. In the case of two neighbouring qubits, the small separation between dots leads to crosstalk effects, whereby the frequency of a qubit is affected by the top gate of the neighbouring dot as well \cite{Cifuentes22023}. In the general case, this leads to the requirement for simultaneous pulsing on both gates to address one of the qubits without affecting the other.

SiMOS spin qubits are known to exhibit weak spin-orbit coupling, which is further minimised by pointing the static magnetic field along [100]~\cite{tanttu2019}. While this is advantageous to achieve degeneracy between the spins, it also impacts the magnitude of the Stark shifts from the top gates~\cite{Cifuentes2023,tanttu2019}. Fortunately, only a small shift ($<{10}\%$ of $\Omega_{\rm{R}}$) is required for qubit control. In fact, too large shifts would cause the rotating wave approximating (RWA) to break \cite{laucht2016b}, and moreover, would leave the qubits more susceptible to electrical noise. In this device we are able to Stark shift Q2 by $\sim200$\,kHz using gate P2 in \autoref{fig:fig4}(a), compared to a Rabi frequency of 1\,MHz. 

We find that the effect of gate P2 on Q2 is $\sim4$\,times stronger than on Q1. This strong differential Stark shift control is sufficient to achieve good addressability of Q2. We note that the inverse was not true and we did not perform operations using P1. Instead, we perform all single qubit gates in the same dot using P2 only and perform SWAP operations to address both qubits. In \autoref{fig:fig4}(c-d) individual control of both qubits is shown with positive $y$-rotation on Q1 and negative $y$-rotation on Q2 for different P2 voltage amplitudes and different initial states. The initial state is determined by a microwave pulse applied to the degenerate qubits in the $\ket{\downarrow\uparrow}$ state, similar to what was done in \autoref{fig:fig2}(d-f). In \autoref{fig:fig4}(e-f) the same sequence is used but with positive rotations on both qubits. Together, the data in \autoref{fig:fig4} (c,d) and (e,f) demonstrate universal single-qubit control. 

The single-qubit gate duration is $6\times{T_{\rm{mod}}}$ and the SWAP duration $1\times T_{\rm{mod}}$ with $T_{\rm{mod}}=1.8\,$\textmu{s}. $T_{\rm{mod}}$ can be reduced by increasing $\Omega_{\rm{R}}$ proportionally. The total sequence duration of $>23\,$\textmu{s} in \autoref{fig:fig4} leads to a significantly reduced visibility similar to \autoref{fig:fig3}(d). 

\subsection*{Conclusion}
\label{sec:conc}
We have shown that two degenerate spins can be driven synchronously with a single global field and universally controlled electrically by local electrodes. This represents an alternative control strategy tackling the inevitable problem of frequency crowding in large arrays of spins. The spins are continuously decoupled from environmental noise by driving them on-resonance throughout any computation. We find that the driving is particularly important during entangling gates between spin states that are exposed to dephasing. In summary, this work represents a paradigm shift in qubit control strategies using dressed degenerate spins for noise-robust and scalable universal control.

\twocolumngrid
\section*{References}
\bibliography{bib11}

\subsection*{Methods}
\label{sec:methods}
{\noindent\bfseries{Experimental setup. }}The SiMOS quantum dot device is fabricated using 800\,ppm isotopically purified $^{28}$Si with a gate stack of Al/AlO$_x$ oxide. It is the same device as device A in Ref.~\onlinecite{Tanttu2023} operated in isolated mode, that is, electrically isolated from the nearby electron reservoirs. The electron configuration is (3,3), except for \autoref{fig:fig1}(f) which is (1,3).  The two dots are formed under gates P1 and P2. The gate J1 is used to control the exchange coupling between the spins. A global field is generated with an on-chip ESR antenna. Spin information is read out with a single-electron-transistor (SET) using Pauli spin blockade (PSB).

The experiments are done in an Oxford Kelvinox 400HA dilution refrigerator. DC bias voltages are coming from Stanford Research Systems SIM928 Isolated Voltage Sources. Gate pulse waveforms are generated with a Quantum Machines (QM) OPX+ and combined with DC biases using custom linear bias combiners at room temperature. The SET current is amplified using a room temperature I/V converter (Basel SP983c) and sampled by a QM OPX+. The microwave pulses are generated with a Keysight PSG8267D Vector Signal Generator, with I/Q and pulse modulation waveforms generated from the QM OPX+. The vector magnet is an Oxford instruments MercuryiPS.

The only feedback protocol used in this work is on the SET top gate, monitoring the current I$_{\rm{SET}}$ to maintain maximum sensitivity.

\vspace{1cm}
{\noindent\bfseries{Interleaved measurements. }} The data displayed in \autoref{fig:fig2}(a-c), \autoref{fig:fig2}(d-f), \autoref{fig:fig2}(g-i) and \autoref{fig:fig3}(c-d) are all taken in an interleaved manner including two or three data sets. An interleaved measurement protocol involves acquiring a single data point for each data set before stepping the measurement parameters so that data points are acquired for each data set sequentially. This reduces temporal bias when trying to make a fair comparison between data sets.   

\section*{Data availability}
All data of this study will be made available in an online repository.

\section*{Code availability}
The analysis codes that support the findings of the study are available from the corresponding authors on reasonable request.

\section*{Acknowledgments}
We acknowledge support from the Australian Research Council (FL190100167 and CE170100012), the U.S. Army Research Office (W911NF-23-10092) and the NSW Node of the Australian National Fabrication Facility. The views and conclusions contained in this document are those of the authors and should not be interpreted as representing the official policies, either expressed or implied, of the Army Research Office or the U.S. Government. The U.S. Government is authorised to reproduce and distribute reprints for Government purposes notwithstanding any copyright notation herein. I.H., A.E.S., S.S., M.K.F., J.Y.H., and A.N. acknowledge support from Sydney Quantum Academy.

\section*{Author contributions}
C.H.Y., A.S.D., I.H., A.S. and A.L. conceived the project and experiments. A.E.S. and M.K.F. developed theoretical models. W.H.L. and F.E.H. fabricated the device. K.M.I. prepared and supplied the $^{28}$Si epilayer wafer. S.S., J.Y.H., N.D.S. and T.T. assisted in experiments and with the cryogenic measurement setup. I.H. and A.N. performed the experiments. I.H. and C.H.Y. analysed the data and wrote the manuscript with input from all the authors.

\section*{Competing interests}
A.S.D. is CEO and a director of Diraq Pty Ltd. T.T., N.D.S., W.H.L., F.E.H., A.S., A.L., A.S.D., and C.H.Y. declare equity interest in Diraq Pty Ltd. I.H., A.E.S., A.S., C.H.Y., A.L., and A.S.D. are inventors on a patent related to this work (PCT 2023004469) filed by the University of New South Wales. 

\newpage 
\onecolumngrid

\newpage 
\captionsetup[figure]{name={Extended data Fig.},labelsep=period,justification=raggedright,font=small}
\setcounter{figure}{0}    

\vspace{1cm}

\section*{Extended data}
\begin{figure*}[hbt!]
    \centering
    \includegraphics[width=0.42\textwidth]{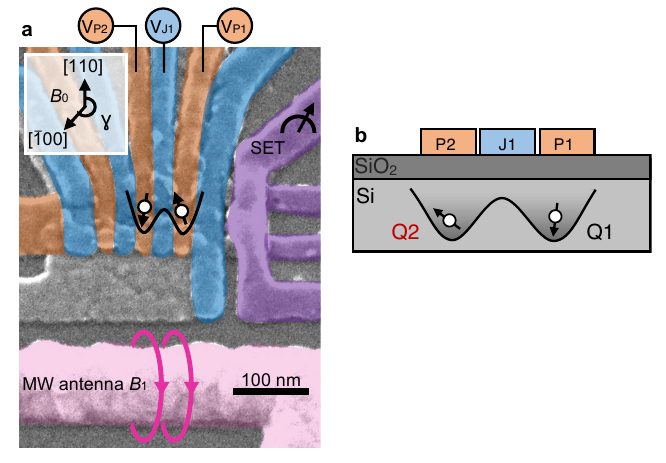}
    \caption{{\bfseries{Device layout.}} (a) False-coloured scanning electron microscope (SEM) image of an identical SiMOS quantum dot device. A double dot is confined under gate P1-P2 with electron configuration (3,3). The gate J1 is used to control the exchange coupling between the spins. A global field is generated with an on-chip ESR antenna. A device cross-section view is given in (b).}
    \label{fig:ext0}
\end{figure*}
\begin{figure*}[hbt!]
    \centering
    \includegraphics[width=0.95\textwidth]{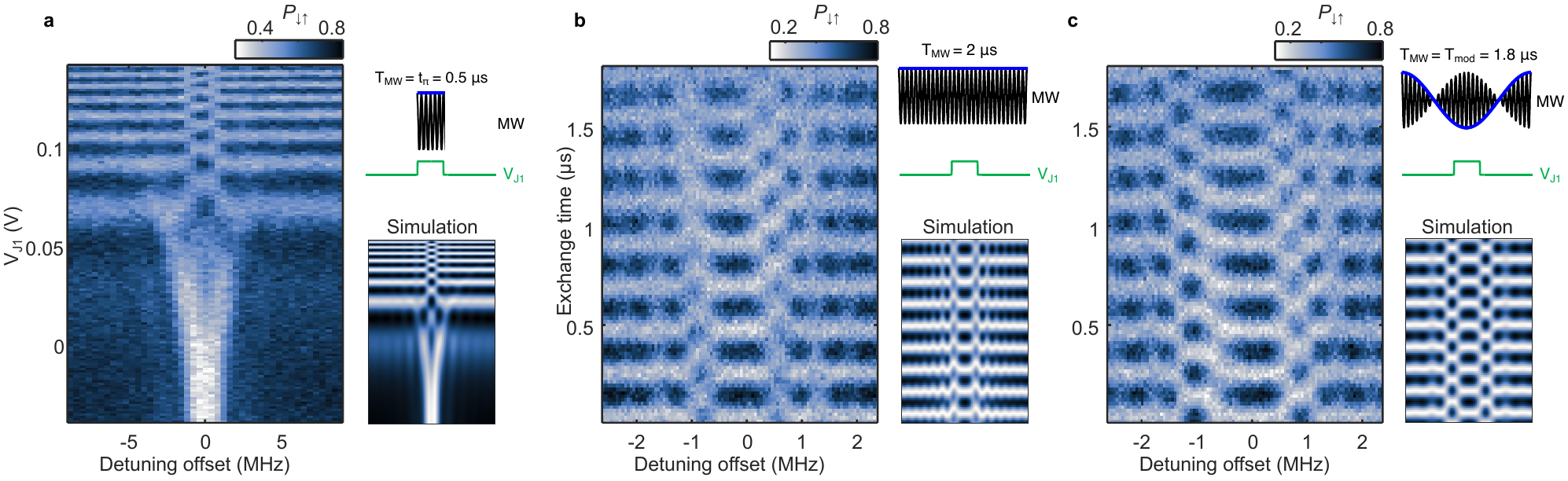}
    \caption{{\bfseries{Larmor frequency robustness of driven SWAP oscillations.}} (a) ESR spectrum as a function of V$_{\rm{J1}}$ where the on-resonance microwave performs a $\pi$-rotation. (b) Square modulation and (c) sinusoidal modulation SWAP oscillation at a fixed V$_{\rm{J1}}$ level for different microwave frequencies and increasing time, acquired in an interleaved manner. The total microwave duration for the square modulation is 2\,\textmu{s} and for the sinusoidal modulation 1.8\,\textmu{s}. Simulations are shown for each sub-panel.}
    \label{fig:ext2}
\end{figure*}
\begin{figure*}[hbt!]
    \centering
    \includegraphics[width=0.8\textwidth]{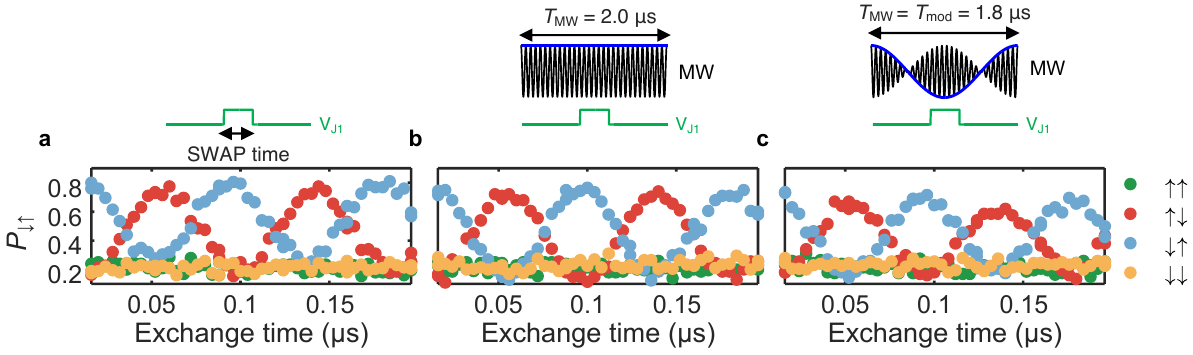}
    \caption{{\bfseries{SWAP oscillation for different initial states.}} SWAP oscillations with initialisations $\{\ket{\downarrow\downarrow},\ket{\downarrow\uparrow},\ket{\uparrow\downarrow},\ket{\uparrow\uparrow}\}$ for (a) no microwave, (b) continuous wave microwave and (c) sinusoidal modulated microwave.}
    \label{fig:ext}
\end{figure*}
\newpage
\captionsetup[figure]{name={Supplementary Fig.},labelsep=period,justification=raggedright,font=small}
\setcounter{figure}{0}   
\section*{Supplementary materials}

\noindent{{\noindent\bfseries{Hamiltonian.}} The Hamiltonian used for all simulations is shown below, where $\hat{\sigma}_{x,y,z}$ and $\hat{\mathbb{I}}$ are the Pauli operators, 
\begin{equation}
    \hat{H}_{\textrm{2Q}}(t) = \Delta\nu_{\textrm{1}}(t)(\hat{\sigma}_z\otimes\hat{\mathbb{I}}) + \Delta\nu_{\textrm{2}}(t)(\hat{\mathbb{I}}\otimes\hat{\sigma}_z) +\Omega_{\textrm{R1}}(t)(\hat{\sigma}_x\otimes\hat{\mathbb{I}}) + \Omega_{\textrm{R2}}(t)(\hat{\mathbb{I}}\otimes\hat{\sigma}_x) + \sum_{i=x,y,z} J(t) \hat{\sigma}_i\otimes\hat{\sigma}_i
    \label{eq:hamiltonian}
\end{equation}
derived from Refs \cite{Seedhouse2021,Hansen2021}. The detuning of the Larmor frequency of qubit 1,2 from the external magnetic field $B_0$ is $\Delta\nu_{\textrm{1,2}}(t)$, and the Rabi frequencies are $\Omega_{\textrm{R1,2}}(t)$ for qubits 1,2. The exchange amplitude is $J(t)$. }

{The driving field determines the time dependency of $\Omega_{\textrm{R1,2}}(t)$. If the field is constant then $\Omega_{\textrm{R1,2}}(t) = \Omega_{\textrm{R1,2}}$. If the driving field has a sinusoidal modulation, then $\Omega_{\textrm{R1,2}}(t)=\Omega_{\textrm{R1,2}}\sqrt{2}\cos(2\pi f_{\textrm{mod}}t)$ \cite{Hansen2021,hansen2022}. Ideally, $\Delta\nu_{\textrm{1,2}}(t)=0$, except when performing single qubit operations. }

{From the Hamiltonian, it can be seen that when $\Delta\nu_{\textrm{1,2}}(t)\approx0$ and all other parameters are greater than 0, the eigenstates of the system are $\ket{S(1,1)}$, $\ket{T_0(1,1)}$, $\ket{\downarrow\downarrow}$ and $\ket{\uparrow\uparrow}$. This means that oscillations between $\ket{\downarrow\uparrow}$ and $\ket{\uparrow\downarrow}$ are natural, giving rise to Heisenberg SWAP oscillations, as discussed in Refs.~\onlinecite{Meunier2011,Seedhouse2021,Loss1998,Petta2005,Simmons2019,Petit2022}. 
In this work, we use the $\ket{S(1,1)}$ notation to refer to the singlet states in the (3,3) charge configurations, as the first two electrons in each dot form a closed-spin shell and can be disregarded\cite{Yang2020}. Similarly, $\ket{S(2,0)}$ refers to singlet states in the (4,2) charge configurations.}

\vspace{1cm}
{\noindent\bfseries{Optimal sinusoidal modulation.}} The sinusoidal microwave modulation is configured according to Ref.~\onlinecite{Hansen2021} (referred to as the SMART protocol) where the Rabi frequency and the modulation period follows 
\begin{equation}
    \Omega_{\rm{R}}T_{\rm{mod}}=\sqrt{2}/j_i.
\end{equation}
Here, $j_i$ is the $i$-th Bessel root of zeroth order. Using the first root $j_1 = 2.404826$ we find the optimal modulation period to be $T_{\rm{mod}}=1.8$\,\textmu{s} when $\Omega_{\rm{R}}\sim1$\,MHz. By driving the spin qubits with this specific waveform, second order noise cancellation can be achieved.

\vspace{1cm}
\noindent{{\noindent\bfseries{Readout. }} In this work, the readout method is based on Pauli spin blockade \cite{Seedhouse2021_0}. This is achieved by tuning the energy levels of the two neighbouring quantum dots, each containing a spin $1/2$ qubit, such that only the $\ket{S(1,1)}$ state tunnels into the $\ket{S(2,0)}$ state, while no other spin state can -- following the Pauli exclusion principle. In our case, due to the coupling of $\ket{\downarrow\uparrow}$ with $\ket{S(1,1)}$, only the $\ket{\downarrow\uparrow}$ state transitions into $\ket{S(2,0)}$, which we call adiabatic ST readout. Alternatively, if there is strong coupling between $\ket{T_{0}}$ and $\ket{S(1,1)}$, set by the Zeeman energy difference, the readout results in the parity readout regime where only $\ket{T_{0}}$ and $\ket{S(1,1)}$ transition into $\ket{S(2,0)}$. The charge movement from (1,1) to (2,0) distinguishes between a state that is blockaded and a state that is unblockaded. In this work, we were able to achieve the adiabatic ST and parity regime by changing the Zeeman energy difference at the read point -- the larger the Zeeman energy difference, the more likely $\ket{T_{0}}$ will decay into $\ket{S(2,0)}$, resulting in parity readout.}
\begin{figure*}[hbt!]
    \centering
    \includegraphics[width=0.75\textwidth]{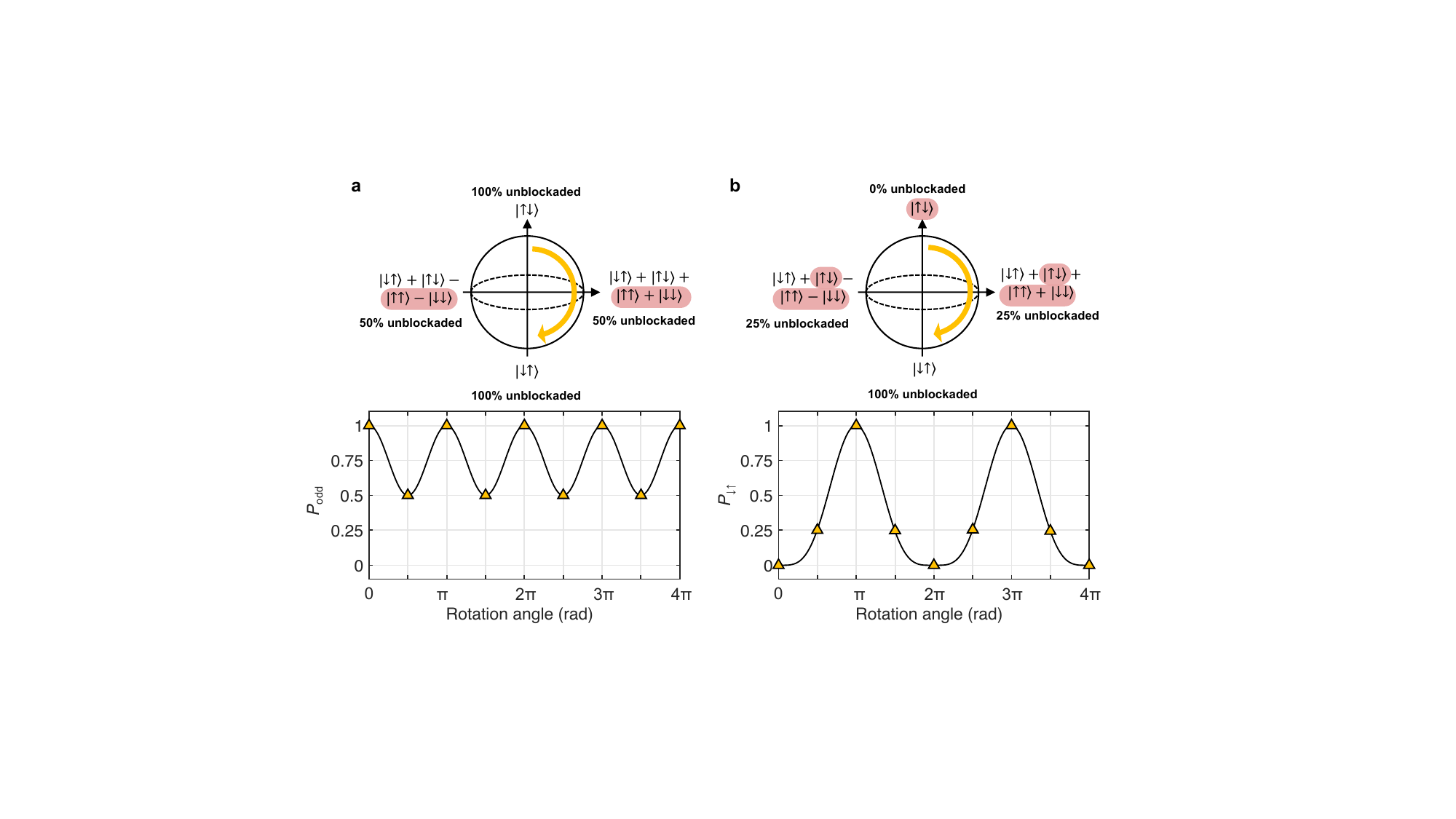}
    \caption{{\bfseries{Rabi probability amplitudes for different readout mechanisms. }}(a) Parity and (b) adiabatic ST readout shown on two superimposed Bloch spheres with the corresponding Rabi probabilities. Here, 1 corresponds to unblockaded and triangle markers show the probabilities of unblockaded at $\pi/2$ intervals.}
    \label{fig:fig5}
\end{figure*}
{To make sense of the readout mechanisms, we discuss the Rabi oscillations observed when both qubits have the same Larmor frequency at the control point. In Supplementary \autoref{fig:fig5}, an initial state $\ket{\uparrow\downarrow}$ is left to evolve on resonance with a driving field, rotating the spins from $\ket{\uparrow\downarrow}$ to $\ket{\downarrow\uparrow}$ as shown by the Bloch spheres. Below the Bloch spheres are the projections of the states into (a) the parity readout basis and (b) the ST basis. In the parity basis, the probability of measuring a charge movement $P_{\textrm{odd}}$ oscillates between 1 (either $\ket{\downarrow\uparrow}$ or $\ket{\uparrow\downarrow}$) and 0.5 (an equal mixture of $\ket{\downarrow\uparrow}$, $\ket{\uparrow\downarrow}$, $\ket{\downarrow\downarrow}$, $\ket{\uparrow\uparrow}$). In the ST basis, the probability of measuring a charge movement $P_{\downarrow\uparrow}$ starts at 0 since $\ket{\uparrow\downarrow}$ is blockaded, then evolves to an equal mixture of all four states, resulting in $P_{\downarrow\uparrow}=0.25$ since only a quarter of the states are blockaded. When the evolution results in $\ket{\downarrow\uparrow}$, the system is fully unblockaded, and $P_{\downarrow\uparrow}=1$. Therefore, adiabatic ST readout exhibits oscillations that decay to 0.75 , while with parity the oscillations decay to $0.5$.

{The ability to distinguish between $\ket{\downarrow\uparrow}$ and $\ket{\uparrow\downarrow}$, i.e. adiabatic ST readout, is useful when observing SWAP operations. To measure SWAP oscillations with parity readout, $\ket{\downarrow\uparrow}$ and $\ket{\uparrow\downarrow}$ are no longer distinguishable, so an additional CNOT gate is needed before readout.}

\newpage 
\vspace{1cm}
{\noindent\bfseries{Adiabatic inversion exchange maps.}}

\autoref{fig:fig1}(d) shows a frequency scan of the qubit energies as a function of the gate voltage ${\textrm{V}_{\textrm{J1}}}$, which controls the exchange interaction. To achieve spin inversion, a chirped microwave pulse (i.e., an pulse whose frequency increases linearly in time) is applied to adiabatically drive spin transitions. The frequency chirp corresponds to a 650\,kHz frequency shift over 33\,\textmu{s}. 

The system is initialized in a $\ket{\downarrow\uparrow}$ state. However, after ${\textrm{V}_{\textrm{J1}}}$ is ramped to the control point, there is a wait time of 100\,\textmu{s} before the adiabatic microwave chirped pulse is applied during which the qubits are left idle and undergo exchange oscillations [see Supplementary \autoref{fig:figexchange}(a)]. In the case where ${\textrm{V}_{\textrm{J1}}}$ is sufficiently high and exchange is large compared to the Larmor frequency difference ${(J>\delta E_{\textrm{Z}})}$, the initialized $\ket{\downarrow\uparrow}$ state undergoes SWAP oscillations. Consequently, at high ${\textrm{V}_{\textrm{J1}}}$ the spin state before the adiabatic microwave chirped pulse becomes a 50/50 mixture of  $\ket{\downarrow\uparrow}$ and $\ket{\uparrow\downarrow}$ or, equivalently, of $\ket{S(1,1)}$ and $\ket{T_{0}}$.

When the adiabatic microwave pulse is applied, we believe three types of transitions can be driven: a first branch that decreases exponentially in frequency corresponds to $\ket{S(1,1)}$ states transitioning into $\ket{T_{-}}$ [light orange arrow in Supplementary \autoref{fig:figexchange}(c)]; a second branch that increases exponentially in frequency corresponds to $\ket{S(1,1)}$ states transitioning into $\ket{T_{+}}$ [dark orange arrow in Supplementary \autoref{fig:figexchange}(c)]; finally, a third branch that remains constant in frequency in the middle of the map corresponds to $\ket{T_{0}}$ states transitioning into $\ket{T_{-}}$ and $\ket{T_{+}}$ (see Supplementary \autoref{fig:figexchange}(c-d) for the energy diagram corresponding to our model of this system). Due to the nature of the parity readout, only $\ket{T_{-}}$ and $\ket{T_{+}}$ will be blockaded, and any mixture of $\ket{S}$ and $\ket{T_{0}}$ will be unblockaded.

The branches corresponding to $\ket{S(1,1)}$ transitions are present. However, instead of the expected blockaded peak associated with $\ket{T_{0}}$ transitions, we observe a central unblockaded peak in between two blockaded peaks. For \autoref{fig:fig1}(d) it is possible to explain this peak at high exchange in a similar fashion as in the case of low exchange interaction, where both spins are being resonantly driven via a multi-photon process and $\ket{\downarrow\uparrow}$  ($\ket{\uparrow\downarrow}$) turns into $\ket{\uparrow\downarrow}$ ($\ket{\downarrow\uparrow}$), maintaining the even parity. When the microwave frequency is slightly off-resonant, partial adiabatic inversion is achieved, leading to a proportion of the spins being in an even state and resulting in the side fringes. Although it can be phenomenologically explained for \autoref{fig:fig1}(d), the absence of a central branch cannot be explained for \autoref{fig:fig1}(c) at $\gamma=224^{\circ}$, therefore it does not match our theoretical model and evidences our incomplete understanding of the dynamics of degenerate spins in a condition of high exchange.
\begin{figure*}[hbt!]
    \centering
    \includegraphics[width=1\textwidth]{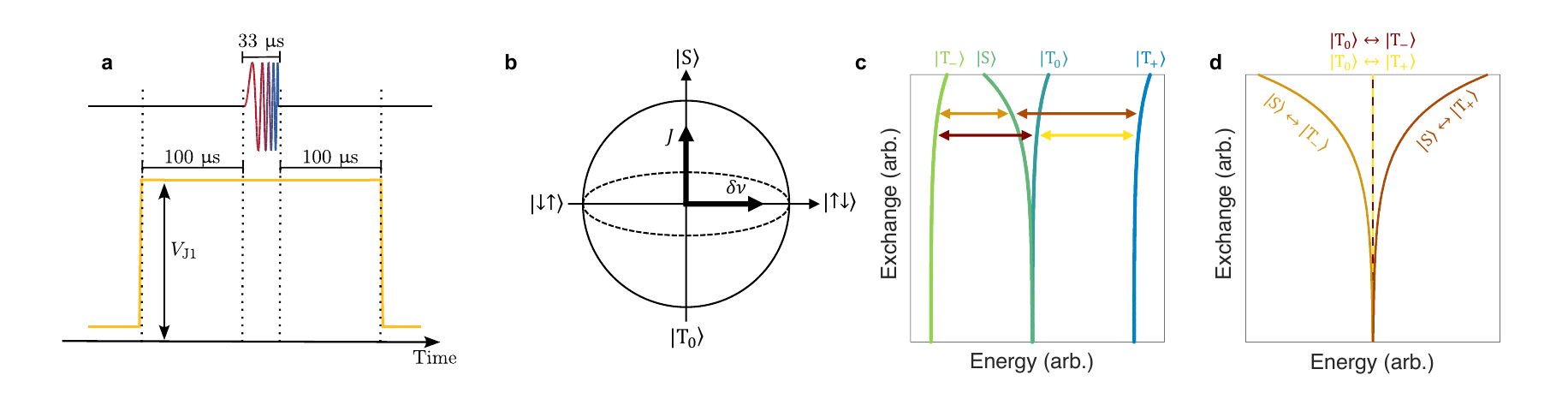}
    \caption{{\bfseries{Adiabatic inversion exchange maps.}} (a) Pulse diagram for the control sequence in the adiabatic scans. The spins are left to freely evolve for $100$\,\textmu{s} before and after the chirped pulse. (b) A ST Bloch sphere indicating the quantisation axis as a function of the ratio of exchange and difference in Larmor frequencies. (c) The energy diagram of a two-qubit system with degenerate Larmor frequencies. The arrows show the different energy splittings in the system. (d) The energy splittings in the two-qubit system as a function of exchange, corresponding to the arrows in (c).}
    \label{fig:figexchange}
\end{figure*}

\newpage
\vspace{1cm}
{\noindent\bfseries{Noise robustness simulations.}} 
During the exchange pulse there is a slight Stark shift from gate J1 that affects the two-qubit system -- consistent with Extended data \autoref{fig:ext2}. The Stark shift will affect any qubit state that is perpendicular to this shift, meaning that any state that is not $\ket{\uparrow}$ or $\ket{\downarrow}$ will be perturbed by the Stark shift. This is seen in \autoref{fig:fig2}(d). For the case of no microwave drive and $\theta=0$, the qubits are not affected by the Stark shift when exchange is on – confirmed by the coherent SWAP oscillation in \autoref{fig:fig2}(d). However, when $\theta=\pi/2$ the state before the exchange is switched on is $\ket{y\bar{y}}$ which is affected by Stark shift. \autoref{fig:fig2}(g) shows a line cut of $\theta=\pi/2$ from (d). The Stark shift has the effect of bringing the qubit system out of sync with the rotating frame frequency, causing beating – consistent with the revival of the SWAP oscillations after 1$\upmu$s. Tuning the device to remove the Stark shift is needed to remove this beating. 

For the case of the driven system with a continuous microwave, when $\theta=\pi/2$ in \autoref{fig:fig2}(e) there is no beating seen. This demonstrates robustness against shifts in the Larmor frequencies. There is a slight reduction in the SWAP oscillation visibility for longer SWAP times when comparing $\theta=\pi/2$ to $\theta=0$, demonstrating some sensitivity to offsets in the Larmor frequencies. Further robustness is seen when the system is driven by the sinusoidal modulated microwave. Comparing the SWAP oscillations in \autoref{fig:fig2}(f) $\theta=\pi/2$ to $\theta=0$, the visibility remains constant. 

To confirm that this observation matches our theoretical interpretation, we simulated the SWAP oscillations for $\theta=\pi/2$ and $\theta=0$, shown in Supplementary \autoref{fig:figswapStarkshift}. The simulation follows the Hamiltonian in \autoref{eq:hamiltonian}, and includes quasi-static Gaussian noise on the amplitude of the Rabi frequency and the Larmor frequency. On top of this, a Stark shift is implemented when the exchange is switched on. The simulation in Supplementary \autoref{fig:figswapStarkshift}(d) shows a revival in the SWAP oscillation for the undriven system when $\theta=\pi/2$, and no effect when $\theta=0$. This is consistent with the experimental data, other than the overall visibility of the SWAP oscillations being affected by more general white noise, or 1/f noise.  Both the continuously driven cases were also simulated with the same offsets in Supplementary \autoref{fig:figswapStarkshift}(b) and (c) for $\theta=0$ and (e) and $\theta=\pi/2$ in (f). Both simulations confirm that neither of the driven cases are strongly affected by the Stark shift. Other noise sources would need to be introduced in the model to further improve the match of the simulation to the experiment.

To demonstrate the improvement in noise robustness for on-resonance driven SWAPs we also simulate the operator fidelity with detuning and microwave noise according to Ref.~\onlinecite{Hansen2021} in Supplementary \autoref{fig:figswap}. Again, we assume quasi-static Gaussian noise. The operator fidelity for the case with no microwave driving, Supplementary \autoref{fig:figswap}(a) and (d), shows that the SWAP operation is highly sensitive to detuning noise compared to the driven cases in (b,c) and (e,f). This agrees with our interpretation of \autoref{fig:fig2}(d-f).
\begin{figure*}[hbt!]
    \centering
    \includegraphics[width=1\textwidth]{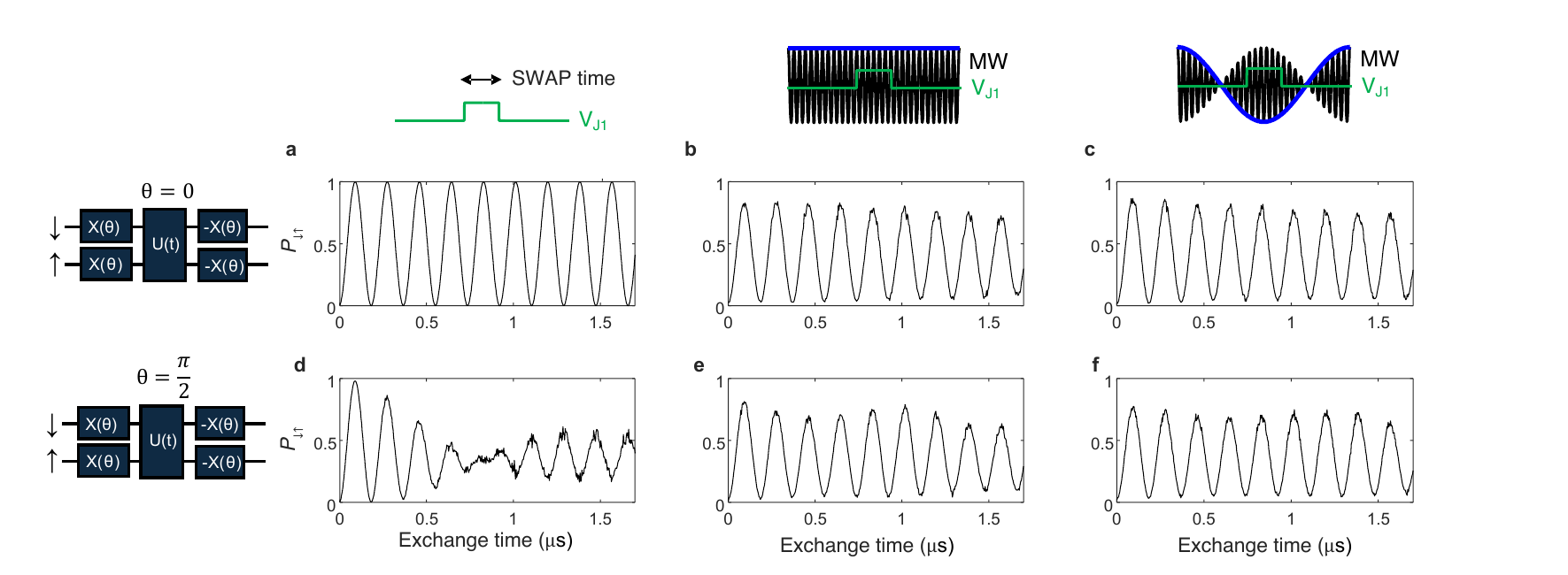}
    \caption{{\bfseries{Simulation of SWAP oscillations with noise.}} Simulations of SWAP oscillations for two different initial states of the qubits for (a) no microwave drive, (b) continuous wave microwave of duration $T_{\rm{MW}}=2.0\,$\textmu{s}, and (c) sinusoidal modulated microwave of duration $T_{\rm{MW}}=T_{\rm{mod}}=1.8\,$\textmu{s} with $\theta=0$. Initialisation with $\theta=\pi/2$ are shown in (d-f). }
    \label{fig:figswapStarkshift}
\end{figure*}
\begin{figure*}[hbt!]
    \centering
    \includegraphics[width=0.7\textwidth]{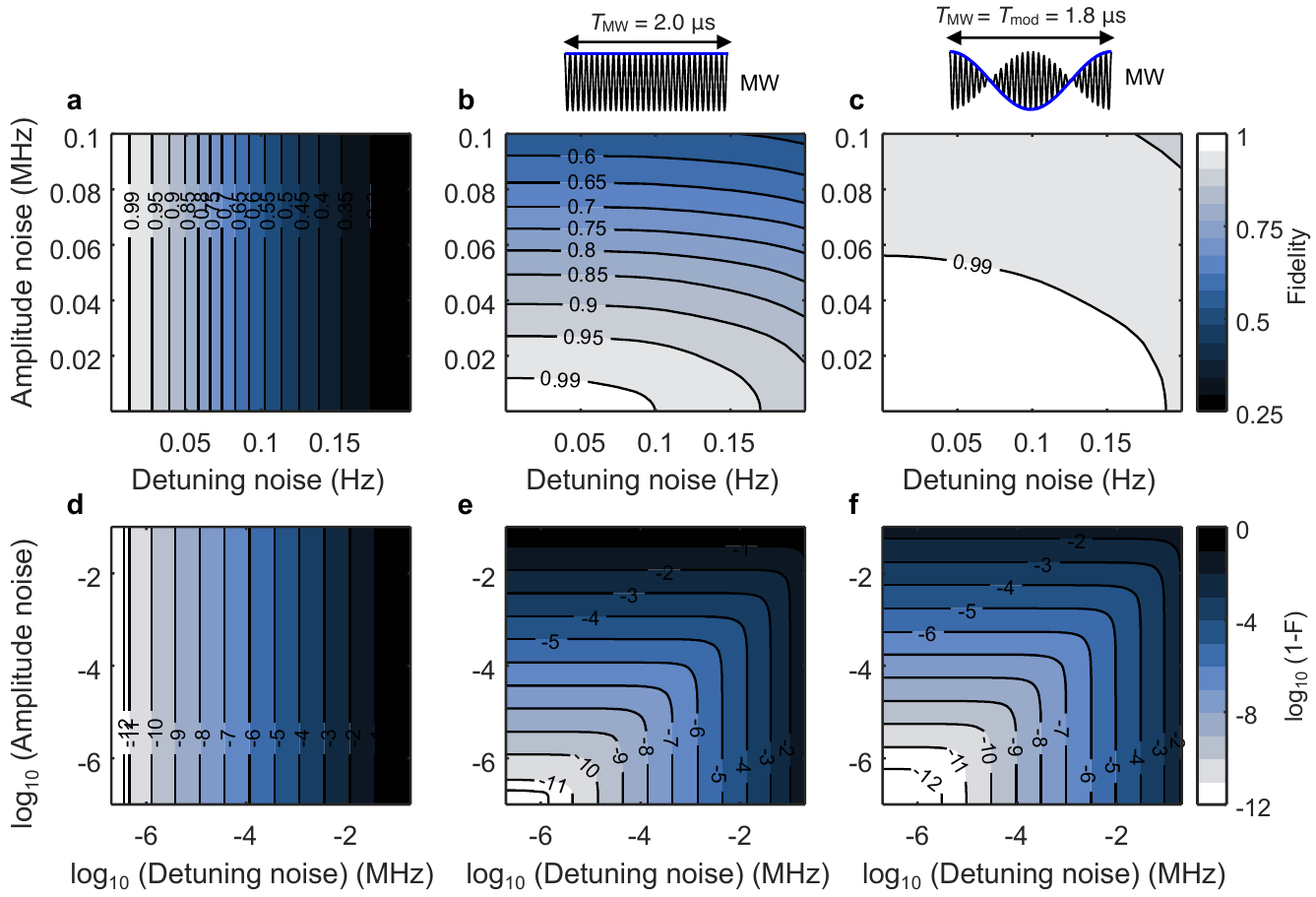}
    \caption{{\bfseries{Simulation of $\sqrt{\rm{SWAP}}$ with noise.}} Operator fidelity of a $\sqrt{\rm{SWAP}}$ gate during (a) no microwave driving, (b) continuous wave microwave and (c) sinusoidal modulated microwave with detuning noise and microwave amplitude noise in linear scale. The same is given in (d-f) in logarithmic scale.}
    \label{fig:figswap}
\end{figure*}

\newpage
\vspace{1cm}
{\noindent\bfseries{Curve fitting of a periodic oscillation.}}
The coherence time of the SWAP oscillations in \autoref{fig:fig2}(g-i) is fitted according to
\begin{equation}
    F(A,b,f,g,\phi,T,t)=A(b+\bigg(\cos{(ft+\phi})\times\cos{(gt})\times e^{-t/T}\bigg)).
\end{equation}
The Ramsey and Hahn echo times in \autoref{fig:fig3}(a-b) are fitted according to
\begin{equation}
    F(A,a,b,c,d,e,f,g,T,t)=A(a-b\bigg(c+d\cos{(ft)}+e\cos{(2ft)}+g\cos{(3ft)}\bigg)\times e^{-t/T}).
\end{equation}
The data in \autoref{fig:fig3}(c-d) is fitted similarly.

\end{document}